\documentclass[sn-basic]{sn-jnl}

\usepackage{amsmath}
\usepackage{amsfonts}
\usepackage{amssymb}
\usepackage{natbib}
\usepackage{colortbl}
\usepackage{here}
\usepackage{physics}
\usepackage{lmodern}
\usepackage{graphicx}
\usepackage[normalem]{ulem}

\jyear{2021}%

\theoremstyle{thmstyleone}%
\theoremstyle{thmstyletwo}%

\theoremstyle{thmstylethree}%

\begin{document}

\title[Rate-induced tipping can trigger plankton blooms]{Rate-induced tipping can trigger plankton blooms}


\author*[1]{\fnm{Anna} \sur{Vanselow}}\email{anna.vanselow@uol.de}

\author[1,2]{\fnm{Lukas} \sur{Halekotte}}\email{lukas.halekotte@uol.de}

\author[3]{\fnm{Pinaki} \sur{Pal}}\email{pinaki.math@gmail.com}

\author[4]{\fnm{Sebastian} \sur{Wieczorek}}\email{sebastian.wieczorek@ucc.ie}

\author[1]{\fnm{Ulrike} \sur{Feudel}}\email{ulrike.feudel@uol.de}

\affil*[1]{\orgdiv{Institute for Chemistry and Biology of the Marine Environment}, \orgname{Carl von Ossietzky University Oldenburg}, \orgaddress{\street{Carl von Ossietzky-Str. 9-11}, \city{Oldenburg}, \postcode{26111}, \country{Germany}}}

\affil[2]{\orgdiv{Institute for the Protection of Terrestrial Infrastructures}, \orgname{German Aerospace Center (DLR)}, \orgaddress{ \city{Sankt Augustin}, \postcode{53757}, \country{Germany}}}

\affil[3]{\orgdiv{Department of Mathematics}, \orgname{ National Institute of Technology}, \orgaddress{ \city{Durgapur}, \postcode{713209}, \state{West Bengal}, \country{India}}}

\affil[4]{\orgdiv{School of Mathematical Sciences}, \orgname{University College Cork}, \orgaddress{\street{Western Road}, \city{Cork}, \postcode{T12XF64}, \country{Ireland}}}


\abstract{Plankton blooms are complex nonlinear phenomena whose occurrence can be described by the two-timescale (fast-slow) phytoplankton–zooplankton model introduced by \cite{Truscott1994}. In their work, they observed that a sufficiently fast rise of the water temperature causes a critical transition from a low phytoplankton concentration to a single outburst: a so-called plankton bloom. However, the dynamical mechanism responsible for the observed transition has not been identified to the present day. Using techniques from geometric singular perturbation theory, we uncover the formerly overlooked rate-sensitive quasithreshold which is given by special trajectories called \textit{canards}. The transition from low to high concentrations occurs when this rate-sensitive  quasithreshold moves past the current state of the plankton system  at some narrow \textit{critical range of warming rates}. In this way, we identify \textit{rate-induced tipping} as the underlying dynamical mechanism. Our findings explain the previously reported transitions to a single plankton bloom, and allow us to predict a new type of transition to a sequence of blooms for higher rates of warming. This could provide a possible mechanism of the observed increased frequency of harmful algal blooms.}




\maketitle

\section{Introduction}
\label{sec:intro}

Marine phytoplankton do not only form the basis of marine food webs and provide approximately half of the global primary production, they also contribute to essential biogeochemical processes in the ocean \citep{Field1998, Falkowski2012}. Accordingly, changes in the presence and timing of elevated phytoplankton concentrations -- called plankton blooms -- may have catastrophic implications for the annual cycle of surface-ocean CO$_2$ uptake or higher trophic levels. 
For example, the release of toxic chemicals in the course of a plankton bloom is able to paralyze or even kill some affected marine species~\citep{Amaya2018,Griffith2019}.

In the past, various abiotic and biotic factors have been identified as potential drivers of harmful and non-harmful plankton blooms. Regarding abiotic factors,  the role of light and mixing processes has been emphasized for many decades~\citep{Gran1935, Riley1946, Riley1949, Sverdrup1953}. The certainly most prominent hypothesis regarding the driving factor is the so-called \textit{critical depth hypothesis} which has already been suggested in the middle of the last century \citep{Gran1935, Sverdrup1953} and which can still be found in more  recently published textbooks \citep{Simpson2012}. Its key element constitutes the \textit{critical mixing depth} above which improved upper ocean growth conditions allow for the accumulation of phytoplankton and thus ultimately for the formation of plankton blooms.

Later, the role of biotic factors such as grazing pressure \citep{Behrenfeld2010, Behrenfeld2013}, competition \citep{Chakraborty2014, Busch2019}, viral infection and parasitism \citep{Chambouvet2008,Velo-Suarez2013,Richards2017} or community composition \citep{Lewandowska2015} attracted increasing attention. These studies include the \textit{disturbance-recovery hypothesis} proposed by \cite{Behrenfeld2010}. The hypothesis presumes that blooms are initiated by recurring physical processes that disrupt the balance between phytoplankton reproduction and grazer consumption. According to the work of Behrenfeld and colleagues, this imbalance is caused by the annual deepening of the mixed layer which ‘dilutes’ the grazing pressure on the phytoplankton which, in turn, allows for its significant accumulation \citep{Behrenfeld2010, Behrenfeld2014}.

However, the phytoplankton reproduction-grazing imbalance
could have many other causes. Rising temperatures and nutrient concentration or increasing light availability together with a reduction of the grazing pressure can enhance phytoplankton growth \citep{Winder2012, Lewandowska2014, Hjerne2019, Trombetta2019, Sommer2011}. In order to fully understand the mechanism underlying the emergence of a plankton bloom, it is not only important to identify the environmental conditions that induce the imbalance, but also to describe how the system escapes from its stable equilibrium state in which the phytoplankton and zooplankton coexist at low densities.
Interestingly, the magnitude of environmental change may not be the one decisive factor for the onset of a plankton bloom, which could explain why the full explanation of this phenomenon has been elusive. The rate of change of environmental conditions also appears to  play an important role \citep{Pinek2020}.

\begin{figure}[t]
\centering
\includegraphics[scale=0.65]{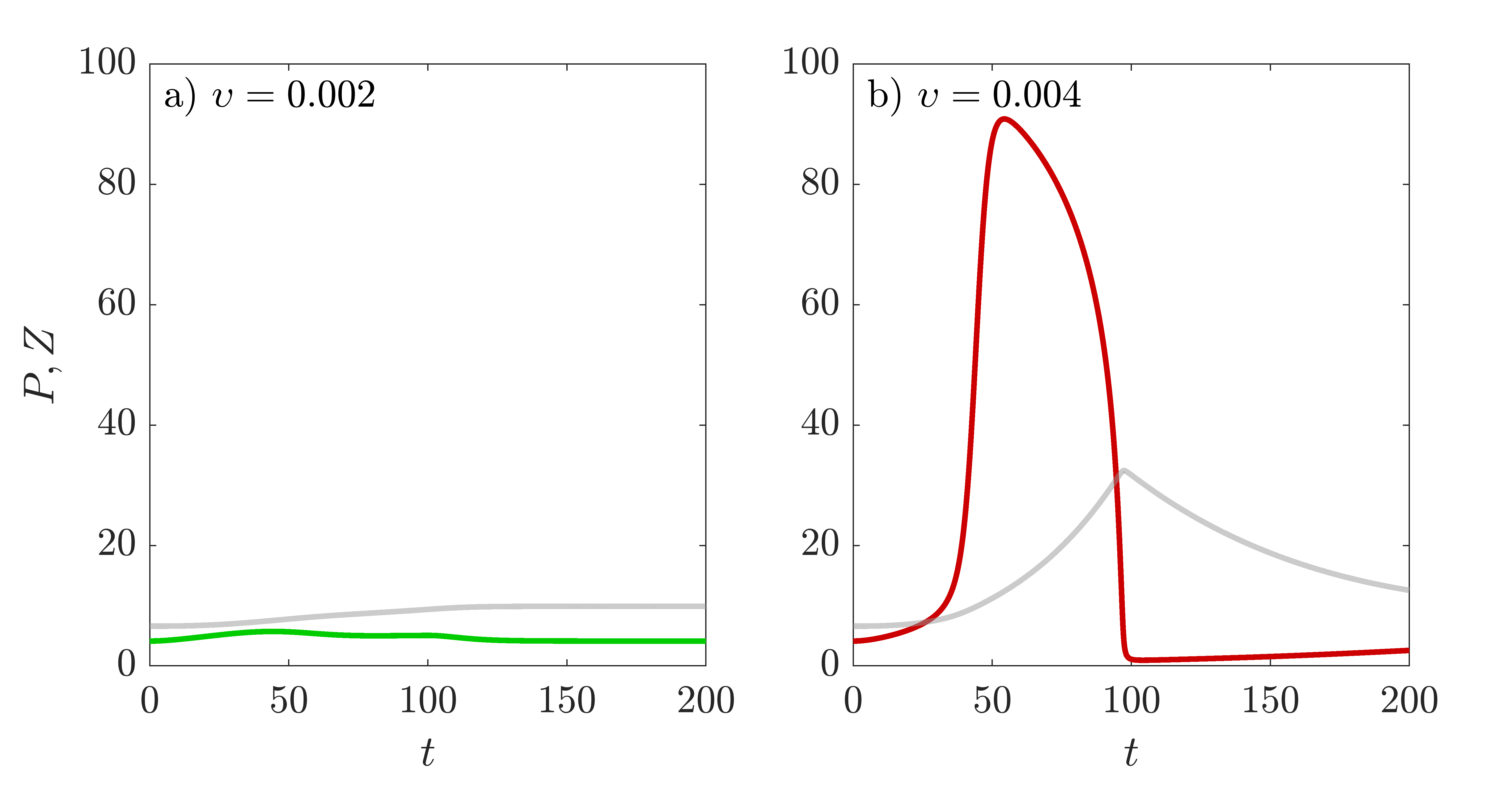}
\caption{The phytoplankton(P) -- zooplankton(Z) model developed by \cite{Truscott1994} is sensitive to rates of environmental change $\upsilon$. It changes its dynamics from the stationary coexistence of $P$ and $Z$ at low densities. (a) to the sudden formation of a phytoplankton bloom (b) if solely the growth rate of the phytoplankton increases faster than the rate $\upsilon \approx 0.003$ day$^{-1}$ (as response to increasing temperatures). The phytoplankton density is displayed in green and red while the zooplankton density is shown in gray. }
\label{fig:intro}
\end{figure}

The indications of rate dependence come from a study in which the Helgoland Roads long-term data series are analyzed \citep{Freund2006}.  The study reveals that the bloom onset is correlated with  rapid increases of temperature, and not so much with the temperature itself. Furthermore, the speed of temperature increase plays a decisive role in the formation of plankton blooms in the theoretical work of \cite{Truscott1994}. Inspired by the idea of an excitable medium capable of mimicking fast growing harmful plankton blooms (red tides), they formulate a plankton model which consists of a fast evolving phytoplankton population that is controlled by a much slower reproducing zooplankton population. Depending on the speed of temperature increase, their time-scaled 
model shows two disparate transient behaviors: (i) balance of phytoplankton and zooplankton at low plankton densities (Fig. \ref{fig:intro}a) or (ii) the formation of a single plankton bloom (Fig. \ref{fig:intro}b). They further ascertain that both behaviors are delimited by a specific value of the speed of temperature increase ($\upsilon \approx 0.003$ in Fig. \ref{fig:intro}).
However, they were not able to identify the {\em rate-sensitive quasithreshold} that separates  these two very different transient responses to temperature increase. In other words, the trigger mechanism of the plankton bloom has not been identified.

Mathematically, the model of \cite{Truscott1994} consists of one fast variable, the phytoplankton, and two slow variables, the zooplankton and the  time-varying environmental condition which changes at a given rate. In such two-timescale systems, exceptional solutions called \textit{canards} are typical \citep{Benoit1981,Benoit1983,Dumortier1996, Szmolyan2001}, and form boundaries between different dynamical regimes \citep{Wieczorek2011,Wechselberger2013,Perryman2014}. For instance, canards can separate small-amplitude oscillations from relaxation oscillations \citep{Brons2008, Desroches2012}, different types of spiking behavior in neuronal models \citep{Izhikevich2007, Mitry2013}, different states of CO$_2$-concentration in the atmosphere \citep{Wieczorek2011,osullivan2022}, or disparate transient \citep{Vanselow2019} or asymptotic \citep{OKeeffe2020} dynamics in biological systems characterized by different dominating species. In the latter four examples, similar to the observations of \cite{Truscott1994}, a variation of the rate of environmental change alone can cause a transition from one dynamical regime to another by crossing canard solutions. These critical transitions have been classified as 
\textit{rate-induced tipping points} by \cite{Ashwin2012}. Others identified so-called \textit{folded-saddle} canards as non-obvious thresholds for the onset of rate-induced tipping \citep{Wieczorek2011,Mitry2013,Perryman2014,osullivan2022}. Specifically, as the rate of environmental change increases, the position of the rate-dependent threshold shifts past the current state of the system, giving rise to a large nonlinear response. It is important to note that in these cases there is no classical bifurcation (no loss of stability in the classical autonomous sense), such as the dangerous saddle-node bifurcation, separating different dynamical regimes. Instead, the different dynamical regimes are solely separated by rate-sensitive canard solutions.

In this work, we uncover the rate-sensitive boundary, that is the singular canard solution, in the model of \cite{Truscott1994}. Furthermore, we demonstrate that this boundary gives rise to a quasithreshold for rate-induced tipping, which is the dynamical mechanism responsible for
the observed plankton blooms (Fig. \ref{fig:intro}).
To this end, we start by introducing the phytoplankton-zooplankton model of \cite{Truscott1994} (Sec. \ref{sec:model}). Afterwards, we reproduce
their simulations in which the phytoplankton population is exposed to rapidly increasing temperatures. Then, we study the resulting rate-sensitive dynamics using geometric singular perturbation theory, and reveal the singular canard solution (Sec. \ref{sec:noBifu}).  Moreover, we show that, depending on the location of the singular canard, the model can show more than one sequentially occurring plankton bloom (Sec. \ref{sec:two_blooms}). Next, we relate this singular canard to a family of canards that form a quasithreshold. Finally, we discuss our results (Sec. \ref{sec:dis}). 

\section{The model}
\label{sec:model}
The Truscott-Brindley model \citep{Truscott1994} (in the following the TB-model) 
combines a logistic growth of a phytoplankton population $P$, a zooplankton population $Z$ growing by the Holling-type III functional response \citep{Holling1959,Holling1959a}, and dying with standard linear mortality. \\
\begin{align}
\label{eq:phy}
    \dv{P}{t} &= r\, P\left(1-\frac{P}{K}\right) - R_m\, Z\,\frac{P^2}{\alpha^2+P^2} &:= f(P,Z,r),\\
    \label{eq:zoo}
    \dv{Z}{t} &= \gamma\, R_m\, Z\, \frac{P^2}{\alpha^2+P^2}-\mu\, Z &:=g(P,Z).
\end{align}
In the following, we vary exclusively the growth rate $r$ of the phytoplankton population while the other parameters are kept constant. The constant parameters $K$, $R_m$, $\alpha$, $\gamma$ and $\mu$ represent the carrying capacity of the phytoplankton population, the attack rate, the half-saturation phytoplankton concentration, the conversion efficiency and the mortality rate of zooplankton population (see Table \ref{tab:parameter} for the parameter values).

We will analyze the response of the plankton model
to a gradual temperature rise, which causes a gradual increase in the phytoplankton growth rate $r$. Specifically, we consider the simplest case in which $r$ increases linearly over time at the rate $\upsilon$ as long as $r\in(r_{min},r_{max})$, and remains constant when $r=r_{min}$ or $r=r_{max}$ (Fig. \ref{fig:model}a):

\begin{equation}
\label{eq:growth_rate}
   \dv{r}{t} = 
   \begin{cases}
    \upsilon>0  &  \text{if} \;\, r_{min} < r < r_{max},\\
    0 & \mbox{if}\;\; r= r_{min}\;\; \mbox{or}\;\; r_{max}.
\end{cases}
\end{equation}

\begin{table}[ht]
    \centering
    \begin{tabular}[h]{ll}
parameter & value  \\
\hline
carrying capacity $K$ & 108 $\mu$g N l$^{-1}$ \\
attack rate zooplankton $R_m$ & 0.7 day$^{-1}$\\
half-saturation constant $\alpha$ & 5.7 $\mu$g N l$^{-1}$\\
conversion efficiency $\gamma$ & 0.05 \\
mortality rate zooplankton $\mu$ & 0.012 day$^{-1}$\\
\hline
minimum growth rate $r_{min}$ & 0.2 day$^{-1}$\\
maximum growth rate $r_{max}$ & 0.6 day$^{-1}$\\
\hline
$P(0)=P_0$ & $e_P$ $\mu$g N l$^{-1}$\\
$Z(0)=Z_0$ & $e_Z$ $\mu$g N l$^{-1}$\\
$r(0)=r_0$ & 0.4 day$^{-1}$
\end{tabular}
\caption{Parameter values of the TB-model \eqref{eq:phy}--\eqref{eq:growth_rate} according to \cite{Truscott1994}.}
\label{tab:parameter}
\end{table}

\subsection{Regular and moving equilibria}

When the temperature is fixed, the  TB-model~\eqref{eq:phy}--\eqref{eq:zoo} with a constant $r$ has three stationary solutions (equilibria). In addition to the extinction and the phytoplankton-only equilibria, both of which are unstable, there is a stable coexistence equilibrium in the $(P,Z)$ phase plane (see app. \ref{app:lts} for more details):

\begin{equation}
    \label{eq:static_state}
      e(r) = \left( e_{P}, e_{Z}(r)\right) = \left( \sqrt{\frac{\mu \alpha^2}{\gamma R_m-\mu} },\,\frac{r}{R_m}\left(1-\frac{e_{P}}{K}\right)\bigg(\frac{\alpha^2 + e_{P}^2}{e_{P}}\bigg) \right).
\end{equation}
This equilibrium corresponds to the balance between the phytoplankton $P$ and zooplankton $Z$ populations, and is the starting point for our analysis of plankton blooms. Most importantly, this equilibrium is linearly stable for the chosen parameter settings (Table~\ref{tab:parameter}) and all $r \in (r_{min}, \; r_{max})$.
In other words, $e(r)$ never bifurcates in the classical autonomous sense within the range of phytoplankton growth rates considered here.

When the temperature rises, the growth rate $r$ changes over time, and we need to consider the extended TB-model~\eqref{eq:phy}--\eqref{eq:growth_rate}, where $r$ becomes an additional dynamical variable.  
When $r = r_{max}$ or $r=r_{min}$, the long term behavior of the TB-model is determined by the unique stable equilibrium $e(r=r_{max})$ or $e(r=r_{min})$, respectively. However, 
when $r \in ( r_{min}, r_{max})$, the system has no equilibrium solutions because $\dv{r}{t} =\upsilon>0$. In this range, we will be interested in how the position of the stable coexistence equilibrium $e(\upsilon t)$ changes with time, and how the system evolves relative to this changing $e(\upsilon t)$.
Note that $e(\upsilon t)$ is referred to as
a quasi-static equilibrium \citep{Ashwin2012, Wieczorek2011}, or a moving equilibrium~\citep{Vanselow2019,OKeeffe2020,Wieczorek2022}. The coexistence of the phytoplankton and zooplankton population at low densities is guaranteed when a solution stays near $e(\upsilon t)$ for small enough growth rates $\upsilon > 0$. In this case, the enhancing growth conditions of the phytoplankton are compensated by an increasing grazing pressure (increasing $r$ causes an increase of $e_Z(r)$, Fig.~\ref{fig:intro}a).
However, an interesting instability 
occurs for larger rates of change $\upsilon$. In this case, the top-down control by the zooplankton relaxes since it is not longer able to balance the faster growing phytoplankton. This provokes an imbalance which results in a deviation of the solution of the extended TB-Model \eqref{eq:phy}--\eqref{eq:growth_rate} from the moving equilibrium  $e(\upsilon t)$, manifesting in the formation of a phytoplankton bloom (Fig.~\ref{fig:intro}b).
This instability cannot be explained by the classical autonomous bifurcation theory, and requires an alternative approach.
To fully understand this instability, we recall some basic concepts from the geometric singular perturbation theory for fast-slow systems. 

\subsection{The critical manifold}

Recall that, in the extended TB-model \eqref{eq:phy}--\eqref{eq:growth_rate}, the phytoplankton population evolves on a different time scale than its grazer. This can be demonstrated by transforming the model into a non-dimensional form. To this end, we introduce the non-dimensional state variables $\hat{P}$, $\hat{Z}$ and the non-dimensional time $\hat{t}$ (see app \ref{app:nondim} for more details):

\begin{align}
    \label{eq:Pnd}
    \gamma \dv{\hat{P}}{\hat{t}} &=\beta \hat{P} (1-\hat{P}) - \hat{Z} \frac{\hat{P}^2}{\delta^2 + \hat{P}^2} &:= \hat{f}(\hat{P},\hat{Z},\beta),\\
    \label{eq:Znd}
    \dv{\hat{Z}}{\hat{t}} &= \hat{Z} \left( \frac{\hat{P}^2}{\delta^2 + \hat{P}^2} - \omega\right) &:= \hat{g}( \hat{P},\hat{Z} ),\\
    \dv{\beta}{\hat{t}} &= 
    \label{eq:rnd}
    \begin{cases}
    \upsilon>0  &  \text{if} \;\, \beta_{min} < \beta < \beta_{max},\\
    0 & \mbox{if}\;\; \beta = \beta_{min}\;\; \mbox{or}\;\; \beta_{max}.
\end{cases}
\end{align}

\begin{figure}[t]
\centering
\includegraphics[scale=0.65]{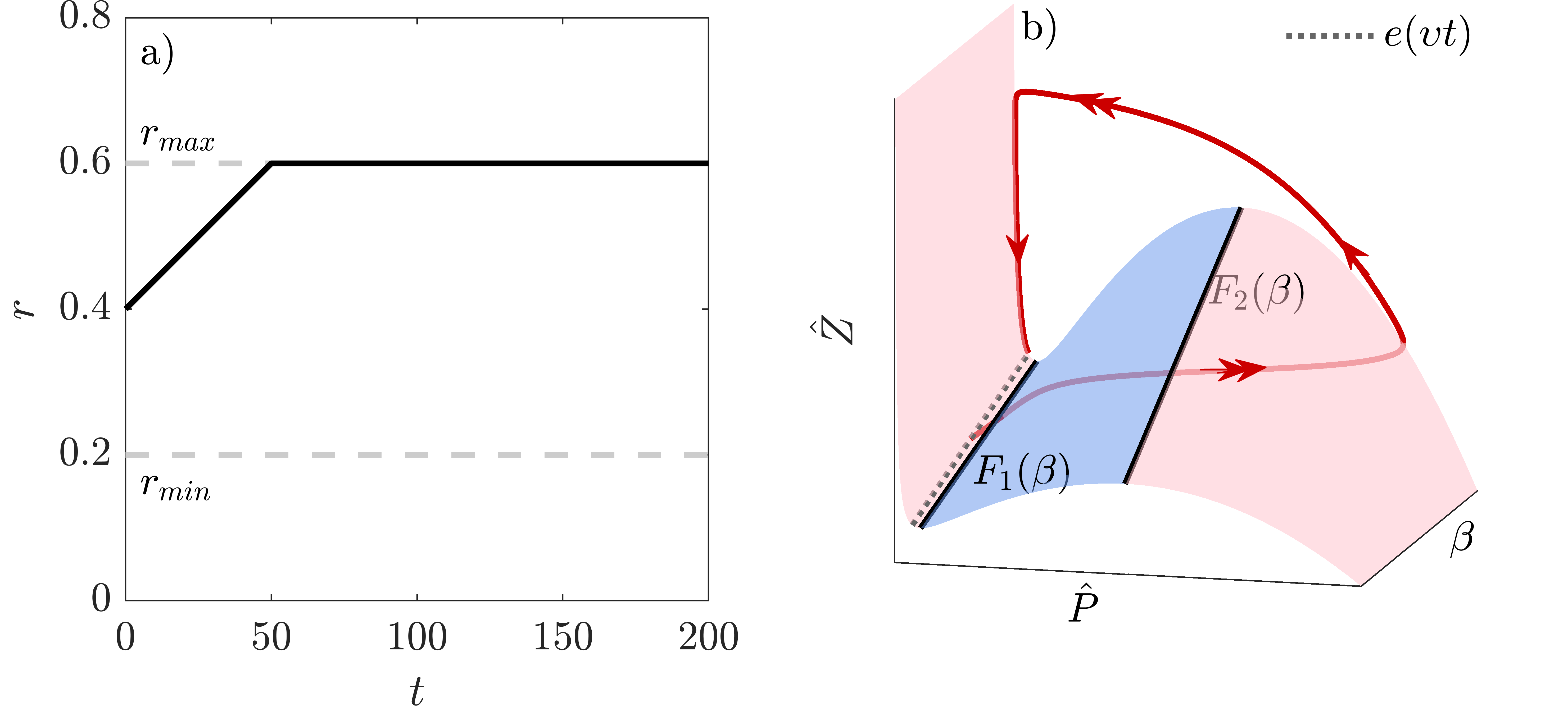}
\caption{(a) The growth rate of the phytoplankton $r$ increases linearly in time at the rate $\upsilon$ between $r_0>r_{min}$ and $r_{max}$ \eqref{eq:growth_rate}. (b) The stable (red), unstable parts (blue) and folds $F_1(\beta)$, $F_2(\beta)$ (black lines) of the critical manifold $\hat{S}$ \eqref{eq:S0construct} organise the slow-fast motion in phase space. Notice that $\hat{S}$ approximates the dynamics in the limit $\gamma \rightarrow 0$. Parameters of red trajectory: (a) $r_0 = 0.4$, $\upsilon = 0.004$, (b,c) $\hat{P}(0) = \hat{P}_0 = \hat{e_P}$, $\hat{Z}(0) = \hat{Z}_0 = \hat{e_{Z}}$, $\beta(0)=\beta_0=0.4$, $\upsilon = 0.004$, $Rm = 0.4$, $\gamma = 0.031$. Other parameters see Tab. \ref{tab:parameter}.  }
\label{fig:model}
\end{figure}

Considering the non-dimensional TB-model \eqref{eq:Pnd}--\eqref{eq:rnd}, it becomes apparent that the time scales separation is determined by the conversion efficiency $\gamma$ of the zooplankton which quantifies its turnover of phytoplankton biomass: the lower the turnover, the slower the zooplankton changes density in time. Following \cite{Truscott1994}, we fix the conversion efficiency to $\gamma = 0.05$. Therefore, the phytoplankton population evolves on the fast time scale $\hat{t}/\gamma$, while the zooplankton density changes on the slow time scale $\hat{t}$.  Additionally, the non-dimensional growth rate 
$\beta$ becomes the second slow variable. The dynamics of such 1-fast 2-slow systems is slow for most of the time (Fig. \ref{fig:model}b, single-headed arrows) with short 
time intervals of fast motion (double-headed arrows). Accordingly, it is reasonable to begin the study of our system by approximating the slow dynamics. 

Taking the limit $\gamma \rightarrow 0$ for the slow time scale $\hat{t}$ gives the reduced problem
\begin{align}
    \dv{\hat{Z}}{\hat{t}} &=  \hat{g}( \hat{P},\hat{Z})
    \label{eq:Zhat}
    \\
    \dv{\beta}{\hat{t}} &= 
    \begin{cases}
    \upsilon>0  &  \text{if} \;\, \beta_{min} < \beta < \beta_{max},\\
    0 & \mbox{if}\;\; \beta = \beta_{min}\;\; \mbox{or}\;\; \beta_{max},
\end{cases}
\end{align}
for the evolution of the slow variables $\hat{Z}$ and $\beta$ in slow time $\hat{t}$ on the so-called critical manifold

\begin{equation}
\label{eq:S0construct}
\hat{S} = \left\lbrace(\hat{P},\hat{Z},\beta) \in \mathbb{R}^3 :  \beta \hat{P} (1-\hat{P}) - \hat{Z} \frac{\hat{P}^2}{\delta^2 + \hat{P}^2}=0, \hat{P} \geq 0,\hat{Z} \geq 0, \beta_{min}<\beta < \beta_{max} \right\rbrace,
\end{equation} 

in the $(\hat{P},\hat{Z},\beta)$ phase space.
Rearranging the condition above with respect to the $\hat{Z}$-coordinate gives the cubic critical manifold formula
\begin{equation}
    \label{eq:S0_hat}
    \hat{Z} = h(\hat{P},\beta) = \beta\left(1-\hat{P}\right)\bigg(\frac{\delta^2 + \hat{P}^2}{\hat{P}}\bigg).
\end{equation}

Besides its approximation of the slow flow (one-headed arrow, Fig. \ref{fig:model}b), the critical manifold organises the fast dynamics as follows: the (red) stable parts of $\hat{S}$ attract the fast flow,  whereas the (blue) unstable part of $\hat{S}$ repels the fast flow. The stable and unstable parts connect at the two folds $F_1(\beta)$ and $F_2(\beta)$ (black lines) of $\hat{S}$. In the following, we display the critical manifold and the trajectories of the TB-model together in several figures. Notice that the critical manifold approximates the dynamics in the limit $\gamma \rightarrow 0$, whereas the trajectories represent solutions of the TB-model for $0 < \gamma \ll 1$. To compare our results, especially the value of $\upsilon\approx 0.003$ around which the transition to a plankton bloom occurs, with the results obtained in the work of \cite{Truscott1994}, we again employ their original formulation of the phytoplankton-zooplankton model \eqref{eq:phy}--\eqref{eq:growth_rate} for the remainder analysis. Naturally, we use the parameter values proposed by \cite{Truscott1994} which follow the reasonable values formulated in \citep{Uye1986,Wake1991} (see Table \ref{tab:parameter}). Furthermore, we assume that the phytoplankton and zooplankton populations are in the quasi-static state $e(\upsilon t)$ \eqref{eq:static_state} at $t=0$ before the growth rate of the phytoplankton $r$ starts to increase linearly in time at the rate $\upsilon$.

\section{Rate-induced tipping triggers plankton bloom}
\label{sec:noBifu}

\begin{figure}[t]
\centering
\includegraphics[scale=0.65]{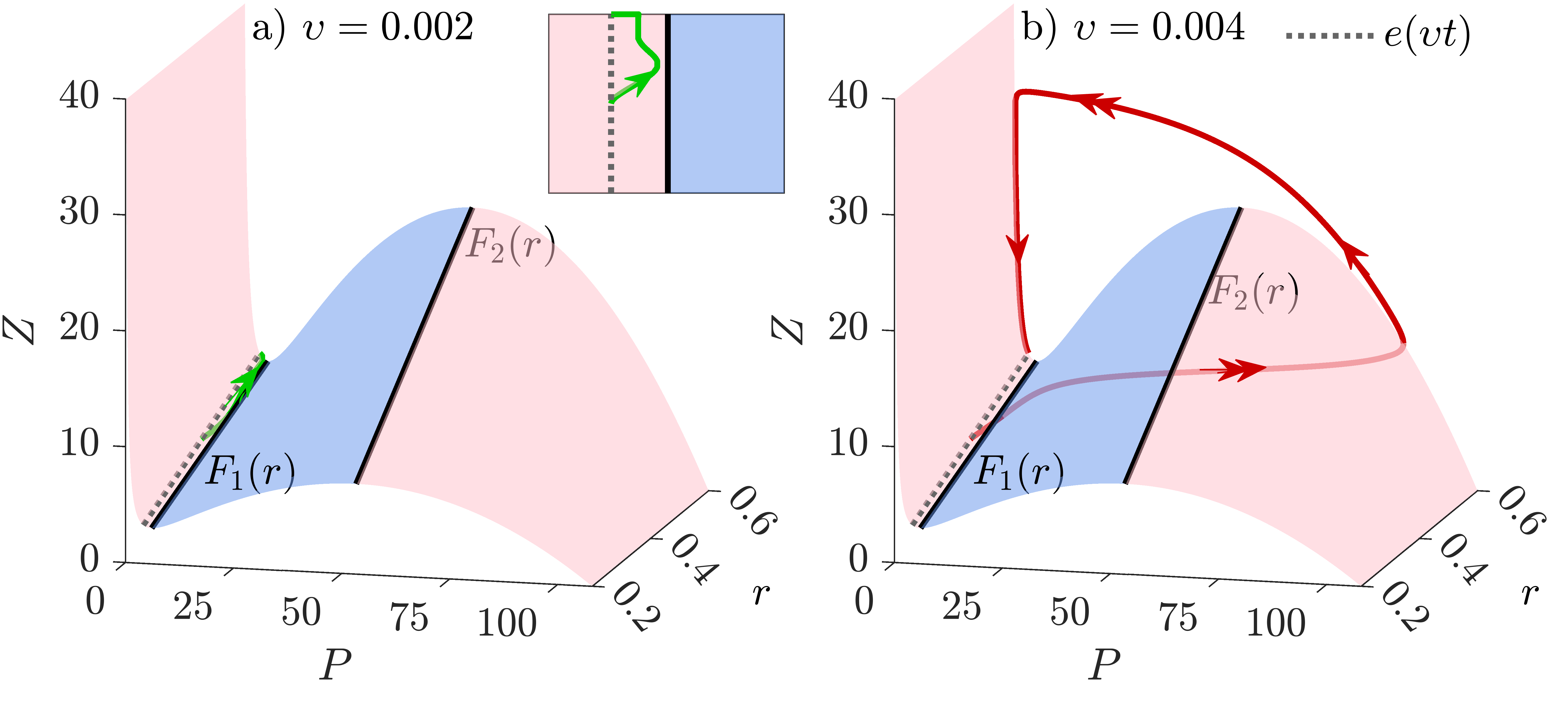}
\caption{Dynamics of the three-dimensional TB-model \eqref{eq:phy}--\eqref{eq:growth_rate} for two  different rates $\upsilon = 0.002$ and $\upsilon = 0.004$. Stable, unstable part and folds $F_1(r)$, $F_2(r)$ of the critical manifold $S$ \eqref{eq:S0} are shown in red, blue and as solid black line. (a) The green trajectory remains close to the quasi-static equilibrium $e(\upsilon t)$ \eqref{eq:static_state} (gray dotted line). (b) The red trajectory leaves the vicinity of $e(\upsilon t)$ close to the fold $F_1(r)$ towards the fast $P$-direction causing the formation of the plankton bloom. $P_0 = e_P$, $Z_0 = e_Z$ and $r = 0.4$. Parameters see Tab. \ref{tab:parameter}. }
\label{fig:move_phase_space}
\end{figure}

If $\upsilon=0.004$, the extended TB-model \eqref{eq:phy}--\eqref{eq:growth_rate} reveals a plankton bloom (Fig. \ref{fig:intro}b, red), but it does not if $\upsilon=0.002$ (Fig. \ref{fig:intro}a, green). Since the difference in $\upsilon$ is the only difference between the two setups (see Table \ref{tab:parameter} for the remaining parameters and the initial conditions), this indicates that, between the rates $\upsilon=0.002$ and $\upsilon=0.004$, the system crosses some quasithreshold that separates both dynamical regimes.
To comprehend what creates the qualitative different behavior of the green and red trajectories (Fig. \ref{fig:intro}), we examine both trajectories in the ($P,Z,r$) phase space (Fig. \ref{fig:move_phase_space}). Since the stable (red), unstable parts (blue) and the folds (black solid lines) of the critical manifold organize the flow in phase space, we add the critical manifold $S$ of the extended TB-model \eqref{eq:phy}--\eqref{eq:growth_rate} to the phase portraits (see app. \ref{sec:S} for the formulation of the critical manifold $S$~\eqref{eq:S0} in terms of the unscaled quantities of \eqref{eq:phy}--\eqref{eq:growth_rate}.)

Starting at $e(\upsilon t)$ \eqref{eq:static_state} at $t=0$, the green trajectory ($\upsilon = 0.002$) reveals the expected behavior: it slowly follows the pathway of $e(\upsilon t)$ (gray dotted line, Fig. \ref{fig:move_phase_space}a) until it settles on the stable long-term state $e(r=r_{max})$. In this case, the slight increase in the zooplankton density, which implies an increasing grazing pressure, is sufficient to balance the growth of the phytoplankton. Consequently, we observe no plankton bloom.
Since the green trajectory remains close to the quasi-static state $e(\upsilon t)$ \eqref{eq:static_state} at all times, we say that the system \textit{tracks} the stable quasi-static state $e(\upsilon t)$.

On the contrary, the red trajectory ($\upsilon = 0.004$) leaves the vicinity of the stable quasi-static state $e(\upsilon t)$ and performs a large excursion before converging to $e(r_{max})$ (Fig. \ref{fig:move_phase_space}b). At first, the red trajectory slowly proceeds towards the fold $F_1(r)$ (black solid line). In the vicinity of $F_1(r)$, its motion changes from slow to fast and remains fast until it reaches the right stable part of $S$. This fast motion away from the quasi-static state $e(\upsilon t)$ entails a rapid increase in the phytoplankton density - the plankton bloom arises. When the trajectory reaches the right stable part of $S$, the plankton bloom possesses its maximal concentration. From here, the red trajectory slowly proceeds towards lower phytoplankton concentrations along the right stable part. Hence, the bloom slowly starts to recede due to an increasing zooplankton concentration controlling it. When the red trajectory approaches the second fold $F_2(r)$, it starts to move fast towards the left stable part: the bloom collapses rapidly due to the still persisting high zooplankton concentration. On the left stable part, it finally converges to the stable long-term state $e(r_{max})$. Since the zooplankton feeds on the phytoplankton on a slower time scale than the phytoplankton reproduces, its evolution in time always lags behind the development of the phytoplankton. For instance, the zooplankton bloom reaches its peak when the phytoplankton density is close to zero (Fig. \ref{fig:move_phase_space}b, Fig. \ref{fig:intro}b).

Since we are particularly interested in what initiates the plankton bloom, we focus on the course of the red trajectory in the following (Fig. \ref{fig:move_phase_space}b). Its switch from slow to fast motion close to the fold $F_1(r)$ of the critical manifold $S$ induces the rapid increase of the plankton population which ends in the bloom. Hence, in order to better understand the initiating mechanism, we examine the evolution of the fast variable $P$ on the critical manifold $S$. To this end, we consider the \textit{reduced system}, which describes the evolution of the slow variables $Z$ and $r$ on the critical manifold $S$. On the basis of the reduced system, we find a three-dimensional system that also captures the evolution of the fast variable $P$ on $S$. Finally, for the sake of simplicity, we lower the complexity of the resulting system by \textit{projecting} its three-dimensional flow on the two-dimensional critical manifold $S$ using $Z = h(P,r)$ (see app. \ref{app:canard} for the transformation of the extended TB-model \eqref{eq:phy}--\eqref{eq:growth_rate} to the projected-reduced system \eqref{fast_flow_in_slow_time}--\eqref{prs2} and also for the detailed formulation of the numerator $\Lambda(P,r,\upsilon)$ and the denominator $\frac{\partial f(P,r)}{\partial P}\Big\vert_S$ in Eq. \eqref{fast_flow_in_slow_time}).
The resulting projected-reduced system is given by:

\begin{align}
\label{fast_flow_in_slow_time}
\dv{P}{t} &= -\frac{\Lambda(P,r,\upsilon)}{\frac{\partial f(P,r)}{\partial P}\Big\vert_S},\\
\label{prs2}
 \dv{r}{t} &=\upsilon.
\end{align}

\begin{figure}[H]
\centering
\includegraphics[scale=0.65]{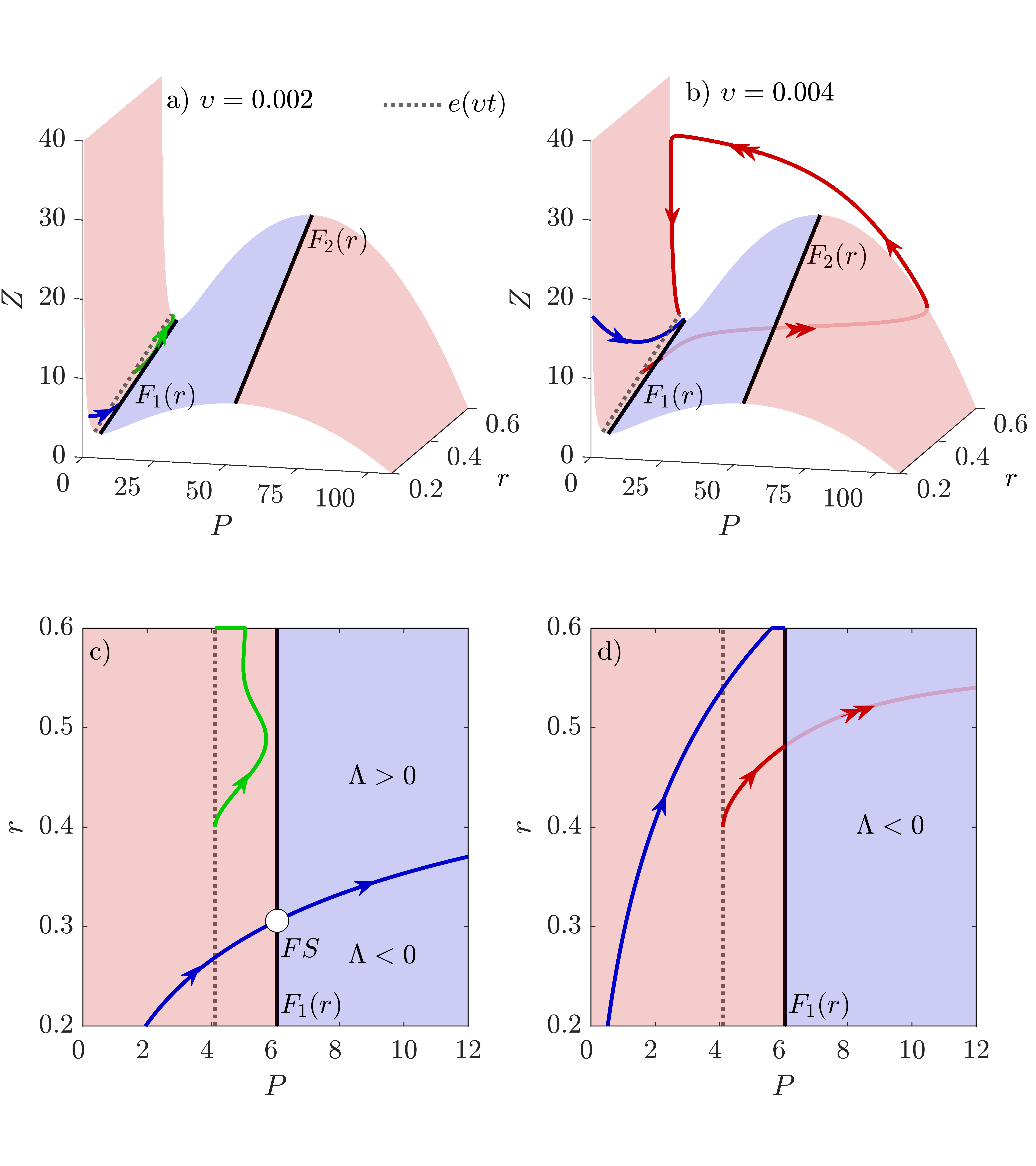}
\caption{(a,c) The green trajectory starts above the canard (blue) where it is repelled by the fold points of $F_1(r)$ ($\Lambda>0$, black line) and returns to $e(\upsilon t)$ -- no bloom emerges. (b,d): The red trajectory starts below the canard where it s attracted by the fold points ($\Lambda<0$), reaches them and moves fast away resulting in the large excursion away from $e(\upsilon t)$ -- the bloom occurs. The canard (blue trajectory) crosses the fold $F_1(r)$ at the folded saddle $FS$ where $\Lambda = 0$. Parameter: $P_0 = e_{P}$, $Z_0 = e_{Z}$, $r_0=0.4$. Other parameters see Tab. \eqref{tab:parameter}.}
\label{fig:noBifu}
\end{figure}

Since the plankton bloom arises when the dynamics of the TB-model \eqref{eq:phy}--\eqref{eq:growth_rate} changes from slow to fast motion close to the fold $F_1(r)$, we start our analysis of the projected-reduced system \eqref{fast_flow_in_slow_time}--\eqref{prs2} 
on the stable part of $S$ near the fold $F_1(r)$ and vary the phytoplankton growth rate $r$. Depending on the sign of the nominator $\Lambda(P_{F1},r,\upsilon)$ which changes with $r$, we can distinguish three different types of trajectories (see Fig. \ref{fig:noBifu} and app. \ref{app:canard} for more details): (i) For growth rates where $\Lambda(P_{F1},r,\upsilon)<0$, (red) trajectories are attracted to the fold $F_1(r)$. However, at the fold $F_1(r)$, the denominator $\frac{\partial f(P_{F1},r,\upsilon)}{\partial P}\Big\vert_S$ becomes zero. Hence, solutions of the projected-reduced system $P(t)$ blow up (go to infinity in finite time $t$) when they reach typical points on the fold $F_1(r)$. In other words, solutions cease to exist within $S$ when they reach typical points on $F_1(r)$ (Fig. \ref{fig:noBifu}bd). 
(ii) For values of $r$ where $\Lambda(P_{F1},r,\upsilon) > 0$, the (green) trajectories never reach $F_1(r)$ because they are repelled from the fold (Fig. \ref{fig:noBifu}ac).
(iii) There can be special points 
(special values of $r$) along the fold $F_1(r)$ at which both the numerator $\Lambda(P_{F1},r,\upsilon)= 0$ and the denominator become zero such that $\dv{P}{t}$ remains finite (see app. \ref{app:canard} for more details). The corresponding (blue) trajectory approaches such point on the fold $F_1(r)$ slowly and is able to cross it with finite speed (Fig. \ref{fig:noBifu}). Afterwards it proceeds slowly along the unstable part of $S$ without being repelled in the fast $P$-direction (Fig \ref{fig:noBifu}c). Hence, this special trajectory - called \textit{singular canard} - combines aspects of both dynamical regimes, i.e. of the green and red trajectory: moving away from the quasi-static state (red, bloom) but slowly (green, no bloom). 
The special fold point $FS = (P_{FS}, r_{FS})$ (Fig. \ref{fig:noBifu}c), at which $\Lambda(P_{F1},r,\upsilon)=0$, is called the {\em folded saddle singularity} \citep{Szmolyan2001} (see app. \ref{app:canard} for the computation of the folded saddle and the canard). 
Since the term $\Lambda(P_{F1},r, \upsilon)$ depends on the rate $\upsilon$ of change of $r$, the position of the folded saddle and thus the position of the singular canard on the critical manifold $S$ also depend on the rate $\upsilon$ (see app. \ref{app:canard}). For this reason, the location of the singular canard changes from the scenario without bloom ($\upsilon = 0.002$, green trajectory) to the scenario with plankton bloom ($\upsilon = 0.004$, red trajectory). 
For $\upsilon = 0.002$, the singular canard is located below the initial condition of the green trajectory. Fold points above the singular canard, respectively above the folded saddle $FS$ ($P_F > P_{FS}$ and $r_F >r_{FS}$), are repelling fold points ($\Lambda > 0$, Fig. \ref{fig:noBifu}c). Consequently, the green trajectory is repelled by the part of the fold $F_1(r)$ above the singular canard and tracks the stable quasi-static state $e(\upsilon t)$: no bloom is formed. 
For $\upsilon = 0.004$, the singular canard is located at higher values of the growth rate $r$. As a consequence, the singular canard is now located above the initial condition of the red trajectory. The fold points below the singular canard are attracting $(\Lambda < 0)$ and therefore, the red trajectory crosses the fold and runs away in the fast direction. As a result, a plankton bloom emerges ($\Lambda < 0$, Fig. \ref{fig:noBifu}d).
Thus, in the limit $\gamma \rightarrow 0$, there is a an isolated {\em critical rate} 
$\upsilon=\upsilon_{crit}$. This critical rate is the value of $\upsilon$ for which the given initial condition lies exactly on the $\upsilon$-dependent singular canard. This singular canard can be thought of as a {\em singular threshold} for rate-induced tipping.

\begin{figure}[t]
\centering
\includegraphics[scale=0.65]{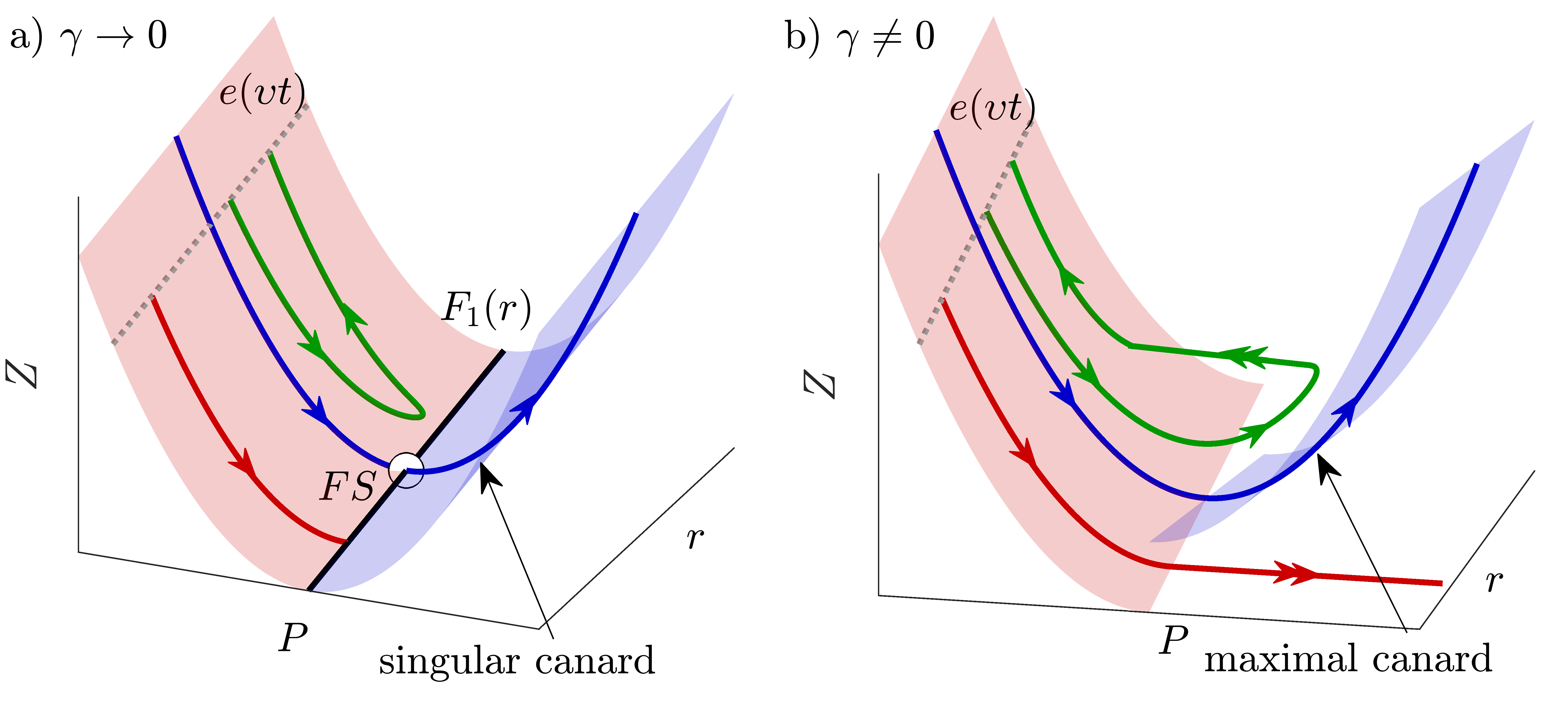}
\caption{(a) Stable (red) and unstable (blue) part of the critical manifold $S$ merge along the fold $F_1(r)$ (black solid line). (Red) trajectories cease to exist at the fold whereas a singular canard (blue trajectory) is able to cross the fold via the point $FS$. (Green) trajectories are repelled by the fold and approach $e(\upsilon t)$ (gray dashed line). (b) Stable and unstable part are perturbed along the position of the fold $F_1(r)$ for $\gamma = 0$. At their intersection point ($FS$ when $\gamma \rightarrow 0$), the maximal canard crosses from stable to unstable part separating (green) trajectories that are repelled by the unstable part from (red) trajectories that leave the stable part in the fast $P$-direction.}
\label{fig:critical_manifold_perturbed_manifold}
\end{figure}

The critical manifold $S$ only approximates the slow dynamics in the limit when the time scale separation $\gamma \rightarrow 0$ \citep{Wechselberger2013}. However, in the full extended TB-model \eqref{eq:phy}--\eqref{eq:growth_rate}, we consider a finite time scale separation between phytoplankton and zooplankton, and, hence,  $0 < \gamma \ll 1$. To evaluate if the singular canard persists in the full extended TB-model (neither reduced nor projected), we need to translate the dynamics in the singular limit ($\gamma \rightarrow 0$) to the full dynamics ($0 < \gamma \ll 1$). In the full system, the critical manifold $S$ is replaced with a nearby slow manifold $S_\gamma$, and the stable and unstable parts of $S_\gamma$ typically split along the fold (Fig. \ref{fig:critical_manifold_perturbed_manifold}b).They only intersect near the point $FS$ (white circle in Fig. \ref{fig:critical_manifold_perturbed_manifold}a) which represents the folded saddle in the singular limit system. This intersection point gives rise to a {\em maximal canard} that crosses from the stable part of the slow manifold into the unstable part of the slow manifold where it stays for as long the unstable part exists (Fig. \ref{fig:critical_manifold_perturbed_manifold}b).
On the lower-$r$ side of the maximal canard, the unstable part is located above the stable part. Hence, (red) trajectories can proceed beneath the unstable part and then run away in the fast direction -- a bloom arises. Conversely, on the higher-$r$ side of  the maximal canard, the unstable part is located below the stable part whereby (green) trajectories proceed above the unstable part and are thus repelled back towards the left stable part -- $P$ and $Z$ remain at low densities.

Therefore, also in the full system ($0<\gamma\ll 1$), there is a boundary  separating  (green) trajectories, reflecting the maintenance of the balance between $P$ and $Z$, from (red) trajectories, that show the formation of a plankton bloom. The difference is that, in contrast to the singular case ($\gamma\to 0$), this boundary is not clear cut. This can be understood as follows. In addition to the maximal canard, which is in a one-to-one correspondence with the singular canard \citep{Wechselberger2013} and follows the unstable part of the slow manifold for the longest time, there are additional canards. These additional canards stay close to the maximal canard for some shorter time, after which they leave the unstable part of the slow manifold towards the left stable part. This whole family of canards is responsible for the transition to a plankton bloom, and is referred to as a {\em quasithreshold} for rate-induced tipping~\citep{Wieczorek2022,osullivan2022}. Crossing such a quasithreshold occurs for  a narrow {\em critical range} of $\upsilon$ rather than at one isolated critical rate $\upsilon=\upsilon_{crit}$.

In summary, we demonstrate that the plankton bloom is solely triggered by rate-induced tipping \citep{Ashwin2012, Wieczorek2011}. An increase of the rate of change $\upsilon$, which describes the speed at which the growth rate of the phytoplankton $r$ increases, changes the position of the maximal canard on the slow manifold $S_\gamma$ in a way that causes the (red) trajectory to leave the region nearby $e(\upsilon t)$, resulting in the formation of a plankton bloom in the full system (Fig. \ref{fig:noBifu}b).

\section{R-tipping can trigger two sequentially occurring plankton blooms}
\label{sec:two_blooms}

\begin{figure}[t]
\centering
\includegraphics[scale=0.65]{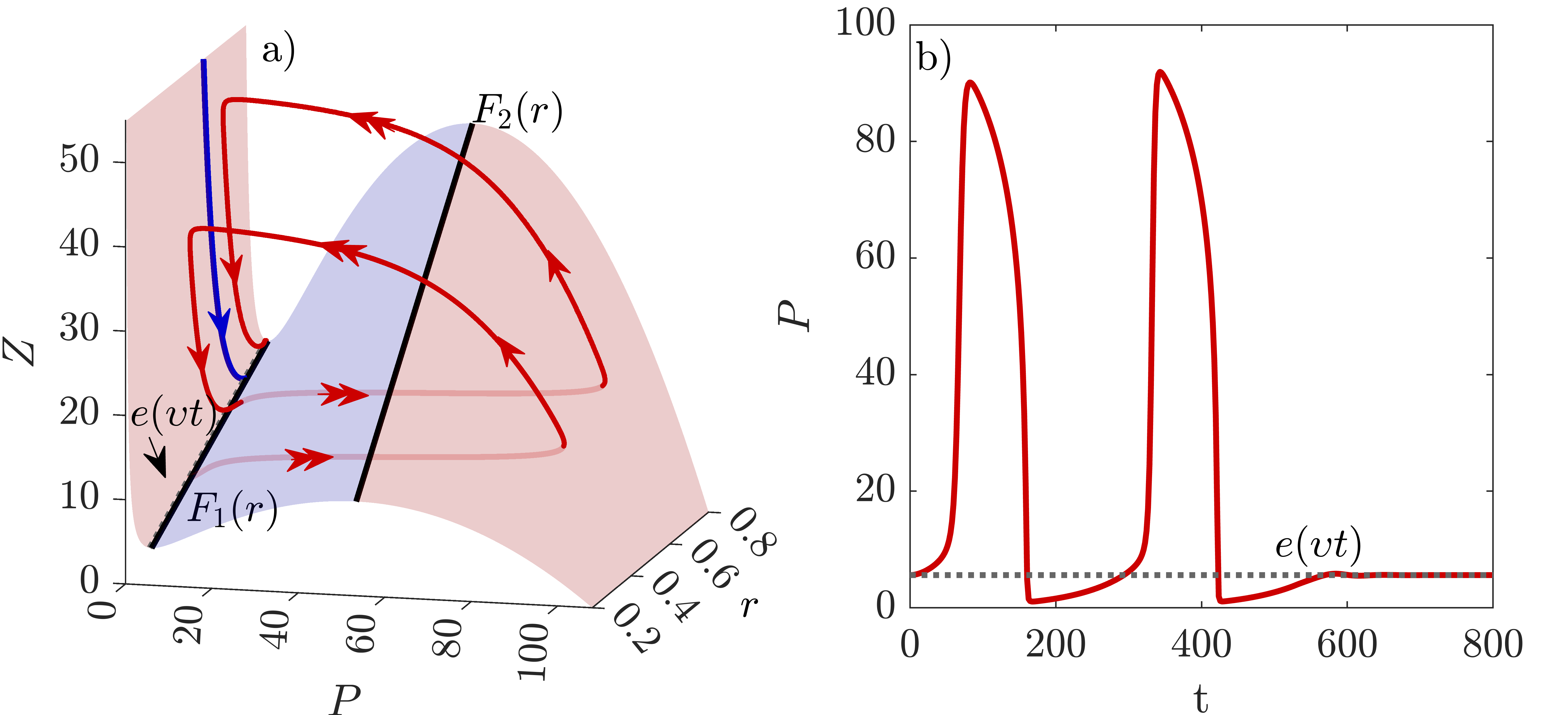}
\caption{Two sequentially recurring plankton blooms occur if the red trajectory reaches the left stable part of the critical manifold $S$ (red) beneath the canard (blue) after performing the first excursion.  Parameter: $P_0 = e_{P}$, $Z_0 = e_{Z}$, $r_0=0.4$, $r_{max} = 0.8$. Other parameters see Tab. \ref{tab:parameter}. }
\label{fig:2blooms}
\end{figure}

In the previous section, we have demonstrated that if initial conditions of the extended TB-model \eqref{eq:phy}--\eqref{eq:growth_rate} are located 
on $S_\gamma$ on the lower-$r$ side of the maximal canard, the model shows  a rate-induced plankton bloom. Clearly, when the red trajectory still finds itself on the lower-$r$ side of the maximal canard following the bloom,
it can exhibit another bloom before it settles on the long-term state $e(r_{max})$ (Fig. \ref{fig:2blooms}). The number of possibly recurring blooms naturally increases with parameter changes that shift the  maximal canard towards higher values of $r$ accompanying a higher maximum growth rate $r_{max}$. For instance, increasing the maximum growth rate $r_{max}$ from $0.6$ to $0.8$ and simultaneously increasing the rate of change $\upsilon$ to $0.006$ triggers two recurring plankton blooms (Fig. \ref{fig:2blooms}). Other parameter changes that promote the occurrence of multiple blooms are an increase of the zooplankton's mortality $\mu$ or a decrease of the zooplankton's attack rate $R_m$ (see app. \ref{app:2blooms} for more details). Obviously, parameter changes reducing the grazing pressure and therefore relaxing the top down control by the zooplankton encourage the formation of more than one bloom.

\section{Discussion}
\label{sec:dis}

In ecology, several factors which are responsible for the occurrence of plankton blooms have already been identified \citep{Behrenfeld2014a,Sommer2012,Lewandowska2015} but still, some mechanisms are not fully understood. Hence, there exists some uncertainty how future climate change will impact the severity and frequency of plankton blooms \citep{Doney2006, Hillebrand2018, Winder2012}. We demonstrate that rate-induced tipping constitutes a possible mechanism explaining the occurrence of plankton blooms. We proceed from the work of \cite{Truscott1994} in which they developed a time scaled phytoplankton($P$, fast variable)-zooplankton($Z$, slow variable) model using ideas from the theory of excitable media. In their work, they studied the response of the model to fast environmental changes that provoke an increase of the phytoplankton's growth rate. Depending on its speed of change, the model reveals two disparate behaviors: conservation of the balance of $P$ and $Z$ (no bloom), or an imbalance of both that manifests in a plankton bloom. Using fast-slow system theory, we uncover the  quasithreshold phenomenon which separates both behaviors in state space: a special trajectory called  maximal canard together with a family of shorter canards. The position of the quasithreshold determines whether solutions remain close to the balance state or leave its vicinity leading to an imbalance that manifests in a plankton bloom. Since the location of the quasithreshold depends on the speed, or the rate, at which the phytoplankton growth rate increases, rate-induced tipping constitutes the mechanism being responsible for the emergence of the plankton bloom. We further demonstrate that decreasing the grazing pressure or broadening the interval in which the growth rate increases allows for multiple recurring plankton blooms. 

In accordance with \cite{Truscott1994}, we assume that the rate-induced bloom is triggered by fast enhancing growth conditions of the phytoplankton due to temperatures increasing at a certain speed. However, other scenarios in which the growth conditions improve gradually at a certain speed are just as conceivable, e.g. due to an increase of the light intensity \citep{Rumyantseva2019,Winder2012} or the nutrient supply \citep{Largier2020,Guseva2020}. Moreover, an imbalance between phytoplankton growth and predator control can be obtained if traits of the grazers, such as attack rate or mortality rate, are affected by rapid environmental changes \citep{Behrenfeld2010, Busch2019}. Consequently, the notion that rate-induced tipping is able to cause plankton blooms does not depend on any particular environmental driver  but rather on its dynamics: As long as any environmental condition changes at a certain speed, rate-induced tipping is a potential trigger mechanism of plankton blooms.  

A second precondition for the occurrence of rate-induced blooms is that phytoplankton reproduces faster than their grazers. This condition is often met since the hourly to daily cell division of phytoplankton \citep{Franks2001} is typically much faster than the reproduction of their grazers which mainly possess generation times from hours to a month \citep{Hirsche2013,Klais2016}. Empirical evidence for the importance of time scale separation in the formation of plankton blooms comes from iron-enrichment experiments carried out in the 90s \citep{Cavender-Bares1999,Cullen1995,Morel1991}. A particularly vivid example was given by the IronEx II experiment \citep{Coale1996}, in which iron enrichment only led to about a doubling of the picoplankton biomass, but up to an 85-fold increase in some species of diatom (phytoplankton). Picoplankton are usually grazed by protists which have often short response time scales—similar to the generation times of the picoplankton. Accordingly, protists can keep the diatoms at low densities. On the contrary, phytoplankton tend to be grazed by metazoan zooplankton, which have relatively long generation times and therefore long response times which allows for the formation of temporary imbalances and thus for the formation of plankton blooms \citep{Franks2001}.

Finally, we want to point out two shortcomings of the current state of the art in ecological modeling concerning the impact of environmental change. We have seen that the observation of rate-induced phenomena, like the formation of plankton blooms, depends on (i) the evolution of the time horizon of environmental change and (ii) the presence of multiple time scales in species' development. (i) Regarding the time scale of environmental change, most studies employ the concept called 'tipping point' which underlies the idea of a catastrophic bifurcation of the involved long-term states if a critical threshold is exceeded \citep{Scheffer2001,Lenton2008,Lenton2020,Kronke2020,Dakos2019}. Bifurcation analysis implies that environmental conditions change infinitely slow. Hence, tipping points are only relevant in cases where it can be assumed that the speed of environmental change is much slower than the intrinsic ecological dynamics. By contrast, rate-induced tipping occurs on time scales comparable to the dynamics of ecosystems and, hence becomes apparent only in the transient dynamics -- the dynamics prior to the long-term dynamics \citep{Hastings2018,Morozov2020}. Hence, the plankton bloom described in this work would be overlooked if we would focus exclusively on the long-term dynamics. \\ 
Concerning the omission of the presence of multiple time scales: if the presence of multiple time scales is neglected, no canard solution can exist and thus, without the existence of any other instability (e.g. a tipping point), no rate-induced phenomena can be observed in the system. Such phenomena are, however, crucial when examining the effect of the speed of environmental change as they reveal critical states which would be missed otherwise. Since present climate change accelerates environmental changes, corresponding studies will be of paramount importance in the future \citep{Walther2002,Parmesan2006,Smith2015}.

In summary, to meet the accelerating speed of environmental change, we suggest three key elements for consideration in future ecological studies. First of all, we believe that it will be of paramount importance to take into account the time horizon of environmental disturbances or changes (which is usually not infinitely slow). Secondly, in order to capture the full impact of fast rates of environmental change, it is essential to explicitly consider the different intrinsic time scales which are present in an ecological system. Thirdly, since critical phenomena -- like rate-induced tipping -- act on time scales comparable to the system dynamics, the evaluation of environmental impacts should not be exclusively based on the long-term response but on the transient dynamics as well.

For certain, the Truscott-Brindley model misses essential physical processes, such as vertical mixing and sinking, as well as other ecological processes, such as viral infection and the diverse composition of interacting communities. However, just like any theoretical model, due to its delightful simplicity it enables the understanding of otherwise unsolvable relationships between ecological actors and their environment. Our findings suggest that plankton communities, which typically involve species evolving on multiple time scales, are potentially prone to environmental disturbances evolving at a certain speed. Hence, we propose to consider multiple time scales in theoretical models of ecosystems, as planktonic food webs, and to examine their transient dynamics.

\section*{Acknowledgement}
This work was supported by the DAAD-DST grant No. 57452783. U.F. acknowledges support by the European Union's Horizon 2020 Research and Innovation program under the Marie Sklodowska-Curie Action Innovative Training Networks Grant Agreement No. 956170 (CriticalEarth). P.P. acknowledges support from the DST-DAAD project (Ref. No.: DST/INT/DAAD/P-15/2019). 






\bigskip
\begin{flushleft}%
Editorial Policies for:

\bigskip\noindent
Springer journals and proceedings: \url{https://www.springer.com/gp/editorial-policies}

\bigskip\noindent
Nature Portfolio journals: \url{https://www.nature.com/nature-research/editorial-policies}

\bigskip\noindent
\textit{Scientific Reports}: \url{https://www.nature.com/srep/journal-policies/editorial-policies}

\bigskip\noindent
BMC journals: \url{https://www.biomedcentral.com/getpublished/editorial-policies}
\end{flushleft}

\begin{appendices}

\section{The Truscott-Brindley model}
\subsection{Non-dimensionalization}
\label{app:nondim}
The phytoplankton($P$)-zooplankton($Z$) model developed by \cite{Truscott1994} is given by the following two equations:

\begin{align}
\label{eq:app_phy}
    \dv{P}{t} &= r P\left(1-\frac{P}{K}\right)-R_m Z\frac{P^2}{\alpha^2+P^2},\\
    \label{eq:app_zoo}
    \dv{Z}{t} &= \gamma R_m Z \frac{P^2}{\alpha^2+P^2}-\mu Z,
\end{align}

with the growth rate of the phytoplankton population $r$ and its carrying capacity $K$. The attack rate of the zooplankton $R_m$, its half-saturation constant $\alpha$, conversion efficiency $\gamma$ and mortality rate $\mu$ complete the model.\\
In the following, we introduce the non-dimensional variables $\hat{P} = \frac{P}{K}$, $\hat{Z} = \frac{Z}{K}$ and the non-dimensional time $\hat{t} = t \cdot R_m$. Replacing $P$, $Z$ and $t$ by their non-dimensional equivalent, we can rewrite the TB-model \eqref{eq:app_phy}--\eqref{eq:app_zoo} as follows:

\begin{align}
\label{eq:ndim1}
      \dv{\hat{P}}{\hat{t}} &=\frac{r}{R_m} \hat{P} (1-\hat{P}) - \hat{Z} \frac{\hat{P}^2}{\left(\frac{\alpha}{K}\right)^2 + \hat{P}^2},\\
    \label{eq:ndim2}
    \dv{\hat{Z}}{\hat{t}} &= \gamma \hat{Z} \left( \frac{\hat{P}^2}{\left(\frac{\alpha}{K}\right)^2 + \hat{P}^2} - \frac{\mu}{\gamma R_m}\right).
\end{align}

Setting $\delta = \frac{\alpha}{K}$, $\beta = \frac{r}{R_m}$ and $\omega = \frac{\mu}{\gamma R_m}$, the non-dimensional TB-model \eqref{eq:ndim1}--\eqref{eq:ndim2} becomes:

\begin{align}
      \dv{\hat{P}}{\hat{t}} &=\beta \hat{P} (1-\hat{P}) - \hat{Z} \frac{\hat{P}^2}{\delta^2 + \hat{P}^2},\\
    \dv{\hat{Z}}{\hat{t}} &= \gamma \hat{Z} \left( \frac{\hat{P}^2}{\delta^2 + \hat{P}^2} - \omega\right).
\end{align}

The parameter $\gamma$ quantifies the separation between the time scale of the phytoplankton and the time scale of the zooplankton. For this reason, we call $\gamma$ the time scale parameter. If $\gamma < 1$, the zooplankton population reproduces slower than the phytoplankton and vice versa. For $\gamma = 1$, it exist no time scale separation.

\subsection{Long-term states and linear stability}
\label{app:lts}
To find the long-term states of the TB-model \eqref{eq:app_phy}--\eqref{eq:app_zoo}, we set the dynamics of the phytoplankton and zooplankton population equal to zero:

\begin{align}
\label{eq:phy0}
    0 &= r P\left(1-\frac{P}{K}\right)-R_m Z\frac{P^2}{\alpha^2+P^2},\\
    \label{eq:zoo0}
    0 &= \gamma R_m Z \frac{P^2}{\alpha^2+P^2}-\mu Z.
\end{align}

Solving Eqs. \eqref{eq:phy0}--\eqref{eq:zoo0} with respect to $P$ and $Z$ leads to the following three long-term states $e_1$--$e_3(r)$ \eqref{eq:e1}--\eqref{eq:e3}

\begin{align}
    \label{eq:e1}
    e_1 &= (e_P,e_Z) = (0,0),\\
    \label{eq:e2}
    e_2 &= (e_P,e_Z) = (K,0),\\
    \label{eq:e3}
    e_3(r) &= (e_P,e_Z(r)) = \left( \sqrt{\frac{\mu \alpha^2}{\gamma R_m-\mu} },\frac{r}{R_m}\left(1-\frac{e_{P}}{K}\right)\bigg(\frac{\alpha^2 + e_{P}^2}{e_{P}}\bigg) \right).
\end{align}

\begin{figure}[H]
\centering
\includegraphics[scale=0.65]{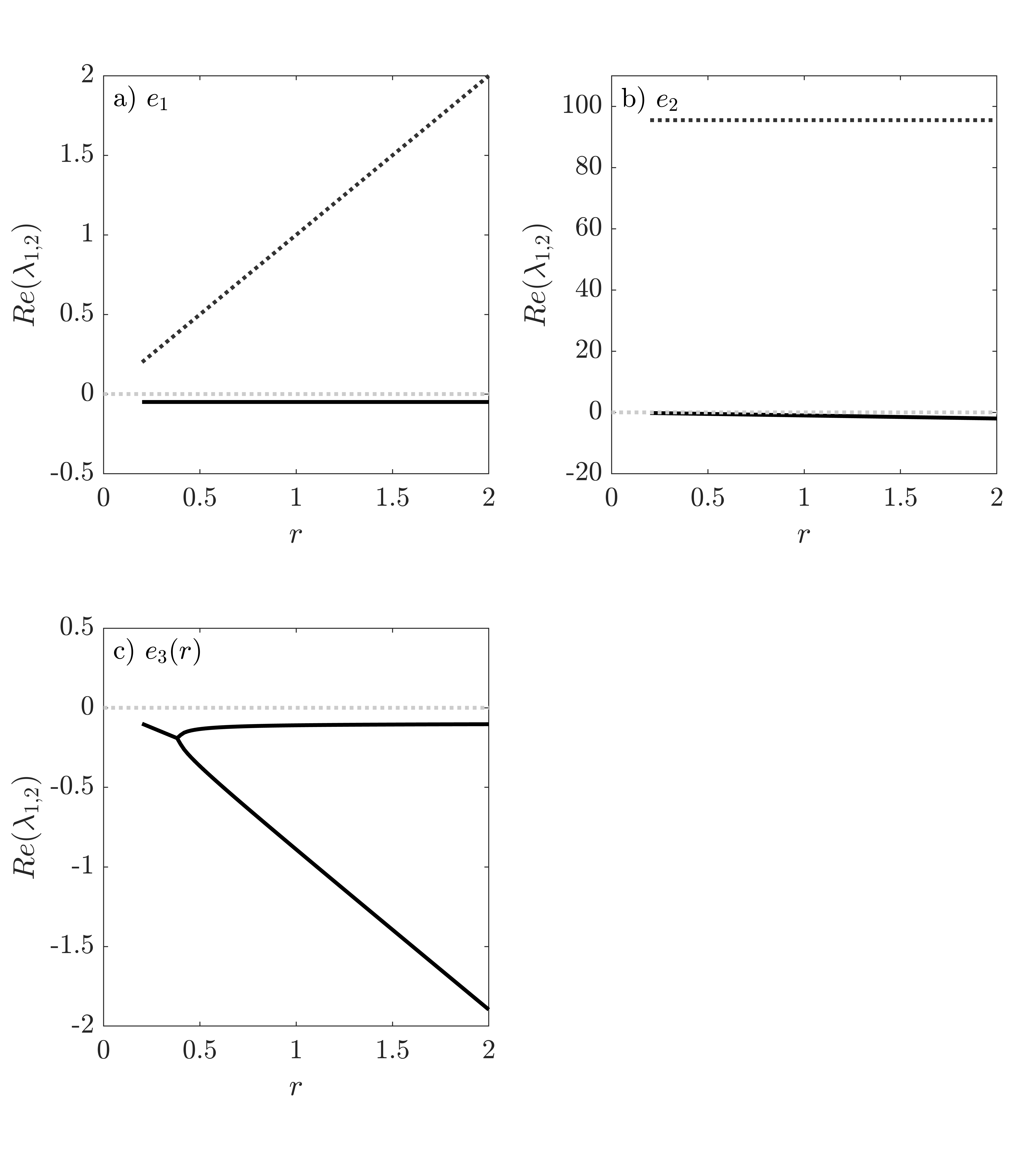}
\caption{Linear stability of the three equilibria $e_1$--$e_3(r)$ \eqref{eq:e1}--\eqref{eq:e3} of the TB-model \eqref{eq:app_phy}--\eqref{eq:app_zoo} depending on the growth rate $r \in [0.2\; 2]$. Solid/dotted line denotes the real parts of the eigenvalues $\lambda_{1,2} < 0$ (stable) respectively $\lambda_{1,2} > 0$ (unstable). For the other parameters see Tab. \ref{tab:parameter}.  }
\label{fig:stability_all}
\end{figure}

The equilibrium $e_1$ describes the situation in which phytoplankton and zooplankton are extinct. In $e_2$, the phytoplankton possesses the maximum density that the environment can 'carry' ($K$) while the zooplankton is extinct. In the third equilibrium $e_3(r)$, phytoplankton and zooplankton coexist at densities unequal zero. Notice that only the equilibrium $e_3(r)$ depends on the growth rate of the phytoplankton $r$. Within the interval $r \in [0.2\; 2]$, it represents the unique stable state of the TB-model (Fig. \ref{fig:stability_all}). Hence for values of the growth rate between $r_{min}$ and $r_{max}$, the phytoplankton and zooplankton coexist in a stable long-term equilibrium when $t \rightarrow \infty$.

\subsection{The critical manifold}
\label{sec:S}

The extended Truscott-Brindley model with time-dependent growth rate $r$ can be written as \citep{Truscott1994}: 
\begin{align}
\label{app:phy_e}
    \gamma\,\dv{P}{t} &= r P\left(1-\frac{P}{K}\right) - R_m\, Z\,\frac{P^2}{\alpha^2+P^2} &:= f(P,Z,r),\\
    \label{app:zoo_e}
    \dv{Z}{t} &= R_m\, Z\, \frac{P^2}{\alpha^2+P^2}-\mu\, Z &:=g(P,Z),\\
    \label{app:growth_rate_e}
   \dv{r}{t} &= 
   \begin{cases}
    \upsilon>0 & \; \text{if} \;\, r_{min} < r < r_{max},\\
    0 & \; \text{if}\;\; r= r_{min}\;\; \mbox{or}\;\; r_{max}.
\end{cases}
\end{align}

The TB-model with time-dependent growth rate $r$ \eqref{app:phy_e}--\eqref{app:growth_rate_e} is determined, as the original TB-model, by a fast and a slow time scale. In fact, the dynamics of such slow-fast systems is primarily slow with short interruption by fast motion. The critical manifold $S$ approximates the slow motion and is therefore a useful tool to obtain a first impression of a part of the full dynamics. 

\begin{equation}
\label{eq:S0}
S = \left\lbrace(P,Z,r) \in \mathbb{R}^3 :   r P\left(1-\frac{P}{K}\right)-R_m Z\frac{P^2}{\alpha^2+P^2}=0, P \geq 0,Z \geq 0, r_{min}\leq r \leq r_{max} \right\rbrace,
\end{equation} 

The critical manifold $S$ \eqref{eq:S0} can be further written as:

\begin{equation}
    \label{app:h}
    Z = h(P,r) = \frac{r}{R_m}\left(1-\frac{P}{K}\right)\bigg(\frac{\alpha^2 + P^2}{P}\bigg),
\end{equation}

with the two-folds $F_{1,2}(r) = (P_{F_{1,2}}, Z_{F_{1,2}})$ given by the following equations

\begin{align}
    \label{eq:fold_P}
    &\frac{2P_{F_{1,2}}^3}{K} - P_{F_{1,2}}^2 + \alpha = 0, \\
    \label{eq:fold_Z}
    &Z = \frac{r}{R_m}\left(1-\frac{P_{F_{1,2}}}{K}\right)\bigg(\frac{\alpha^2 + P_{F_{1,2}}^2}{P_{F_{1,2}}}\bigg).
\end{align}

\section{The canard trajectory}
\label{app:canard}

When studying the slow-fast dynamics of the extended TB-model \eqref{app:phy}--\eqref{app:r} with time-dependent growth rate $r$ for different rates $\upsilon$, we find a plankton bloom when the growth rate $r$ increases faster than $\upsilon \approx 0.003$ (Fig. \ref{fig:intro}, see sec. \ref{sec:noBifu} for more details)

\begin{align}
     \label{app:phy}
    \gamma\dv{P}{t} &= r P\left(1-\frac{P}{K}\right)-R_m Z\frac{P^2}{\alpha^2+P^2} &:= f(P,Z,r)\\
    \label{app:zoo}
    \dv{Z}{t} &= R_m Z \frac{P^2}{\alpha^2+P^2}-\mu Z &:= g(P,Z)\\
    \label{app:r}
    \dv{r}{t} &= \upsilon > 0.
\end{align}

To simplify notations, we use from now on the following abbreviations. Note that $q$ is denoting $f$ and $g$ and $k$ stands for $P$, $Z$ and $r$.

\begin{align}
\label{eq:not1}
    q^S(P,r) &= q(P,Z,r)\big\vert_{Z = h(P,r)},\\
    \label{eq:not2}
    q^S_k(P,r) &= \frac{\partial q(P,Z,r)}{\partial k}\bigg\vert_{Z=h(P,r)}.
\end{align}

The bloom forms when (red) trajectories cross the fold $F_1(r)$ of the critical manifold $S$ and run away in the fast $P$-direction (Fig. \ref{fig:noBifu}). For this reason, we study the fast flow close to the fold $F_1(r)$ on the critical manifold $S$. To study the dynamics on $S$, we set the fast dynamics $\dv{P}{t}$ equal to zero ($\gamma \rightarrow 0 $) which gives the \textit{reduced system}:

\begin{align}
     \label{app:constrain}
    0 &= f(P,Z,r),\\
    \dv{Z}{t} &=  g(P,Z),\\
    \dv{r}{t} &= \upsilon > 0. 
\end{align}

Differentiating the algebraic constraint \eqref{app:constrain} with respect to the slow time $t$ leads to:

\begin{align}
    0 &= \dv{}{t} f(P,Z,r),\\
    0 &=  f_P(P,Z,r)\cdot \dv{P}{t} + f_Z(P,Z,r) \; g(P,Z) + f_{r}(P,Z,r) \upsilon,\\
    \label{app:fast_on_S}
    \dv{P}{t} &= -\frac{f_Z(P,Z,r)\; g(P,Z) + f_{r_t}(P,Z,r) \upsilon}{f_P(P,Z,r)}.
\end{align}

The equation \eqref{app:fast_on_S} describes the fast flow $P$ on the critical manifold $S$. Replacing the constrain Eq. \eqref{app:constrain} by Eq. \eqref{app:fast_on_S}, we obtain:

\begin{align}
    \dv{P}{t} &= -\frac{f_Z(P,Z,r)\; g(P,Z) + f_{r}(P,Z,r) \upsilon}{f_P(P,Z,r)},\\
    \dv{Z}{t} &=  g(P,Z),\\
    \dv{r}{t} &= \upsilon > 0.
\end{align}

Using $Z = h(P,r)$ \eqref{app:h}, we \textit{project} the flow onto the critical manifold $S$. The so-called \textit{projected-reduced system} is given by:

\begin{align}
\label{app:prs}
    \dv{P}{t} &= -\frac{f^S_Z(P,r)\; g^S(P) + f^S_{r}(P,r) \upsilon}{f^S_P(P,r)},\\
      \label{app:prs2}
    \dv{r}{t} &= \upsilon > 0.
\end{align}

With $\Lambda(P,r,\upsilon) = f^S_Z(P,r)\; g^S(P) + f^S_{r}(P,r) \upsilon$ and $f^S_P(P,r) = \frac{\partial f(P,r)}{\partial P}\bigg\vert_{S}$ we can write the projected-reduced system as:

\begin{align}
\label{app:prs_1}
    \dv{P}{t} &= -\frac{\Lambda(P,r,\upsilon)}{\frac{\partial f(P,r)}{\partial P}\bigg\vert_{S}},\\
      \label{app:prs2_1}
    \dv{r}{t} &= \upsilon > 0.
\end{align}

For completeness, we write the projected-reduced system in full terms

\begin{align}
    \dv{P}{t} &= \frac{\frac{-\gamma r R_m P^3 \left( 1-\frac{P}{K}\right)}{\alpha^2+P^2} + \mu r P \left( 1-\frac{P}{K} \right) +\left(P -\frac{P^2}{K}\right) \upsilon}{\left( -r+\frac{2rP}{K} + \frac{2\alpha^2 r\left( 1-\frac{P}{K}\right)}{\alpha^2 + P^2}\right)}\\
    \dv{r}{t} &= \upsilon > 0. 
\end{align}

\begin{figure}[H]
\centering
\includegraphics[scale=0.6]{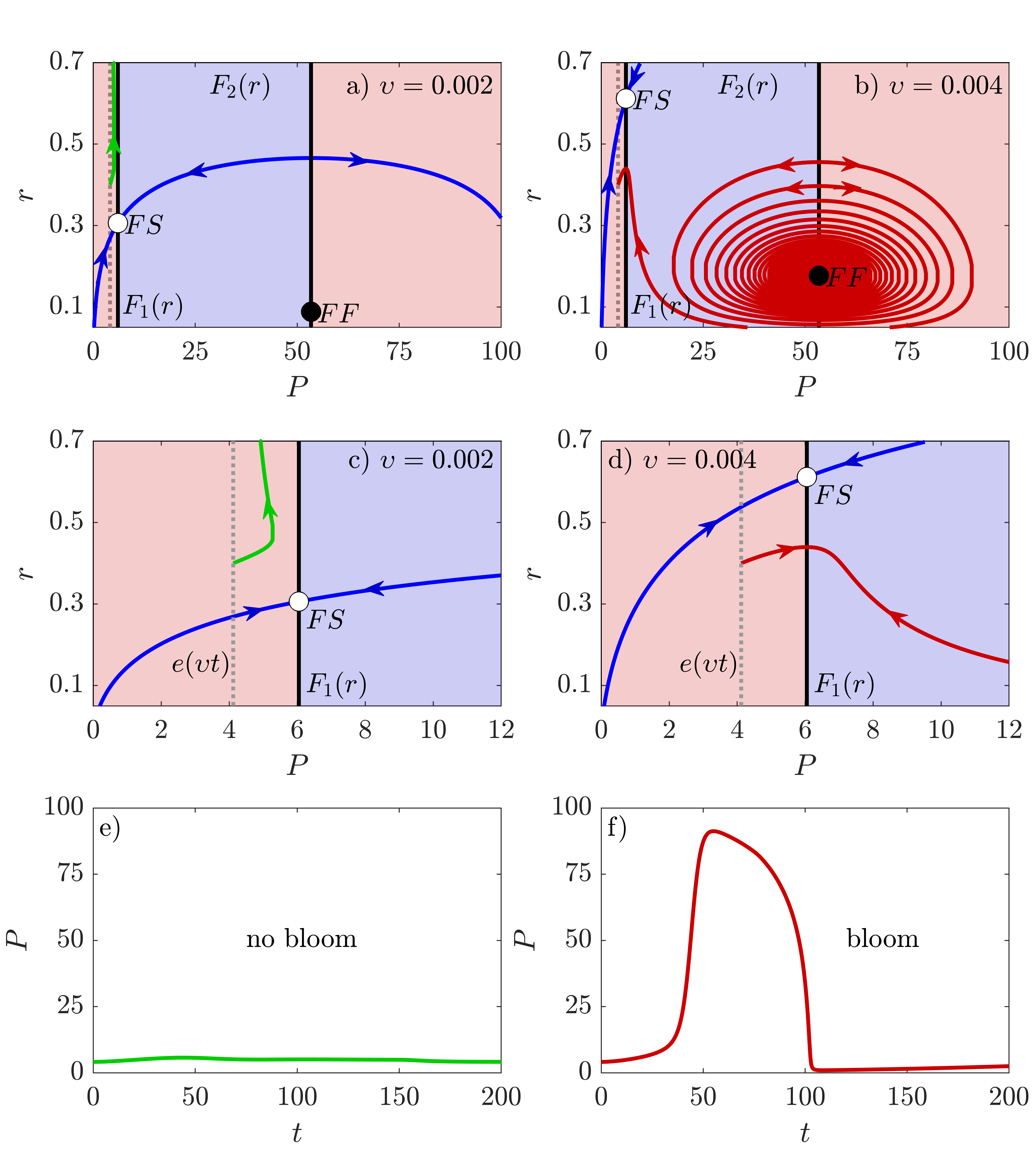}
\caption{Dynamics of the desingularized system \eqref{eq:des1}--\eqref{eq:des2} for two different rates $\upsilon = 0.002$ (a,c,e) and $\upsilon = 0.004$. (b,d,f). The singular canard (blue trajectory) is given by the stable manifolds of the folded saddle $FS$ (white filled circle). It represents the threshold separating (green) tracking from (red) tipping trajectories. (a,c) The green trajectory starts below the singular canard and tracks -- no bloom emerges (e). (b,d) The red trajectory starts below the canard, is attracted by the fold $F_1(p,r)$, crosses it and approaches the stable folded focus $FF$ (black filled circle) -- a plankton bloom occurs (f).    }
\label{fig:des}
\end{figure}

If $f^S_P(P,r) = 0$ \eqref{app:prs}, the flow on the critical manifold $S$ \eqref{eq:S0} is not defined -- it goes to infinity in finite time. Unfortunately, these points for which $f^S_P(P,r) = 0$ are the fold points of the critical manifold $S$ \eqref{eq:S0}. Since the bloom formation occurs close to the fold, we have to enable the analysis near the fold $F_1(r)$. This can be achieved by using the scaling known as \textit{desingularization}:

\begin{equation}
    \label{app:des_scaling}
    t = -\hat{t}\cdot f^S_P(P,r),
\end{equation}

which preserves the direction of time on the stable part of the critical manifold $S$ (red, Fig. \ref{fig:des}), but reverses it on the unstable part (blue, Fig. \ref{fig:des}). The desingularized system is given by:

\begin{align}
    \dv{P}{\hat{t}} &= f_Z^S(P,r)\; g^S(P) + f_{r}^S(P,r) \upsilon, \\
    \dv{r}{\hat{t}} &= -f_P^S(P,r) \upsilon,
\end{align}

respectively in full terms,

\begin{align}
    \label{eq:des1}
    \dv{P}{\hat{t}} &= \frac{-\gamma r R_m P^3 \left( 1-\frac{P}{K}\right)}{\alpha^2+P^2} + \mu r P \left( 1-\frac{P}{K} \right) +\left(P -\frac{P^2}{K}\right) \upsilon,\\
    \label{eq:des2}
    \dv{r}{\hat{t}} &= \left( -1+\frac{2P}{K} + \frac{2\alpha^2 \left( 1-\frac{P}{K}\right)}{\alpha^2 + P^2}\right) \upsilon.
\end{align}

\begin{figure}[H]
\centering
\includegraphics[scale=0.6]{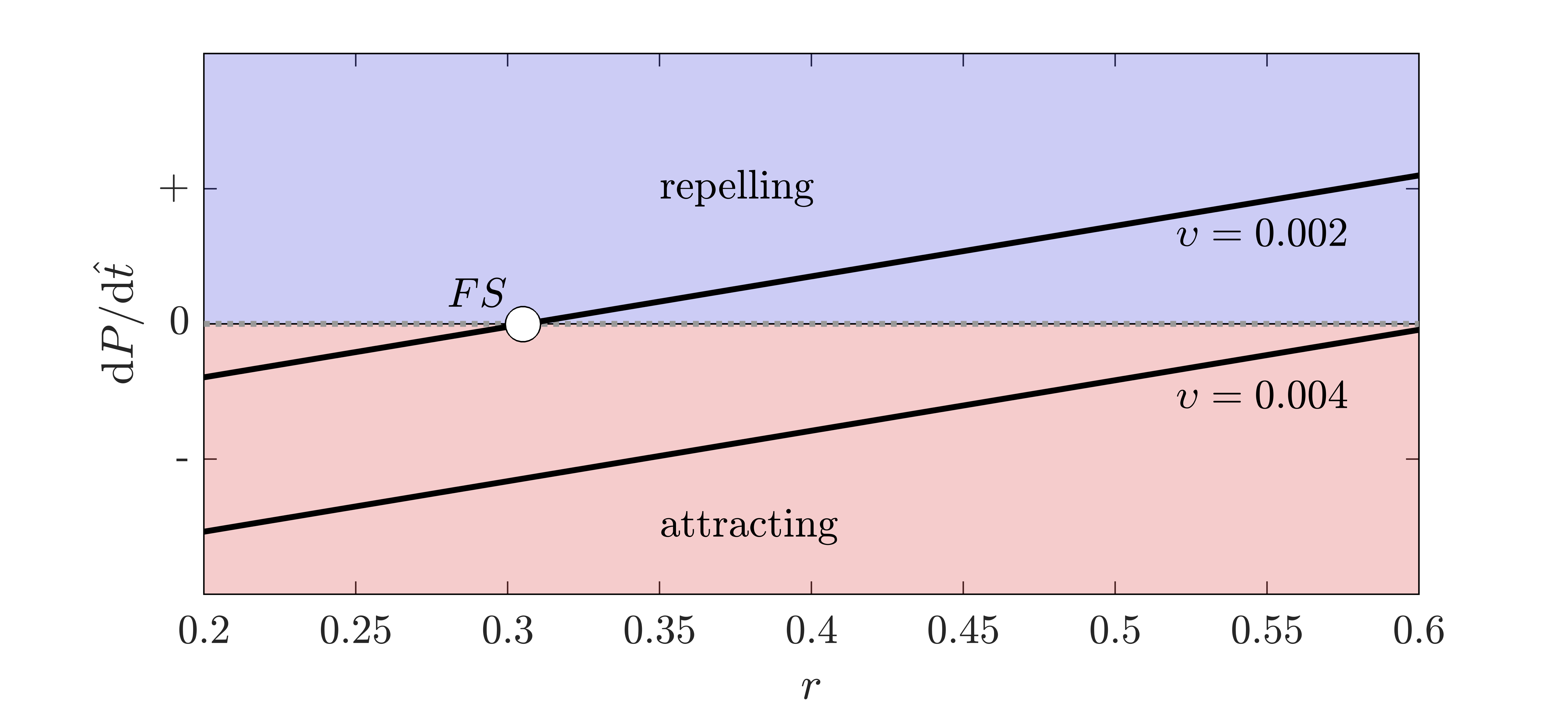}
\caption{The fast flow $\dv{P}{\hat{t}}$ \eqref{eq:des1} (black solid line) on the fold $F_1(r)$ \eqref{eq:fold_P}-\eqref{eq:fold_Z} is shown for two different rates $\upsilon = 0.002$ and $\upsilon = 0.004$. Fold Points with  $\dv{P}{\hat{t}} <0$ attract trajectories (red) while fold points with  $\dv{P}{\hat{t}} >0$ are repelling (green). The fast flow $\dv{P}{\hat{t}}$ remains only finite at the point $FS$. For other parameters see Tab. \ref{tab:parameter}. }
\label{fig:foldpoints}
\end{figure}

Interestingly, setting $\dv{r}{\hat{t}}=0$:

\begin{equation}
    \label{eq:term}
    0 = \left( -1+\frac{2P}{K} + \frac{2\alpha^2 \left( 1-\frac{P}{K}\right)}{\alpha^2 + P^2}\right),
\end{equation}

and rearranging Eq. \eqref{eq:term} with respect to $P$ gives the $P_{F_1}$-values of the fold $F_1(r)$. When we further set $P_{F_1}$ into Eq. \eqref{eq:des1}, we can analyze the fast flow $\dv{P}{\hat{t}}$ at the fold $F_1(r)$ (Fig. \ref{fig:foldpoints}).
For $\upsilon = 0.002$, there exist a special solution at which $\dv{P}{\hat{t}} = 0$ (white circle) within $r \in [0.2\; 0.6]$. This special solution is an equilibrium of the desingularized system \eqref{eq:des1}--\eqref{eq:des2} and it marks the boundary between fold points that attract trajectories ($\dv{P}{\hat{t}} < 0$, red) and fold points that repel them ($\dv{P}{\hat{t}} > 0$, blue). Such special points are called \textit{folded singularities} \citep{Szmolyan2001}. For $\upsilon = 0.004$, we found no folded singularity in the interval $r \in [0.2\; 0.6]$. 

Studying the linear stability of the folded singularity at $F_1(1)$ gives one negative eigenvalue ($\lambda<0$) and one positive eigenvalue ($\lambda > 0$). Hence, it is called a \textit{folded saddle singularity} $FS$. 

And indeed, solutions that cross the fold via the folded saddle singularity show a kind of boundary behavior: they can cross the fold with finite speed and move away from the quasi-static state (bloom) but slowly (no bloom). Such trajectories are called (singular) canards (Fig. \ref{fig:des}). In the desingularized system \eqref{eq:des1}--\eqref{eq:des2} exist besides the folded saddle $FS$ ($\lambda_1 < 0 ,\lambda_2 >0 $) folded focus $FF$ (Re($\lambda_{1,2}$) $< 0$) at the second fold $F_2(r)$. In the case of a folded saddle $FS$, the singular canard trajectory is given by its stable manifolds.

\section{More than one recurring bloom -- parameter studies}
\label{app:2blooms}
\begin{figure}[H]
\centering
\includegraphics[scale=0.6]{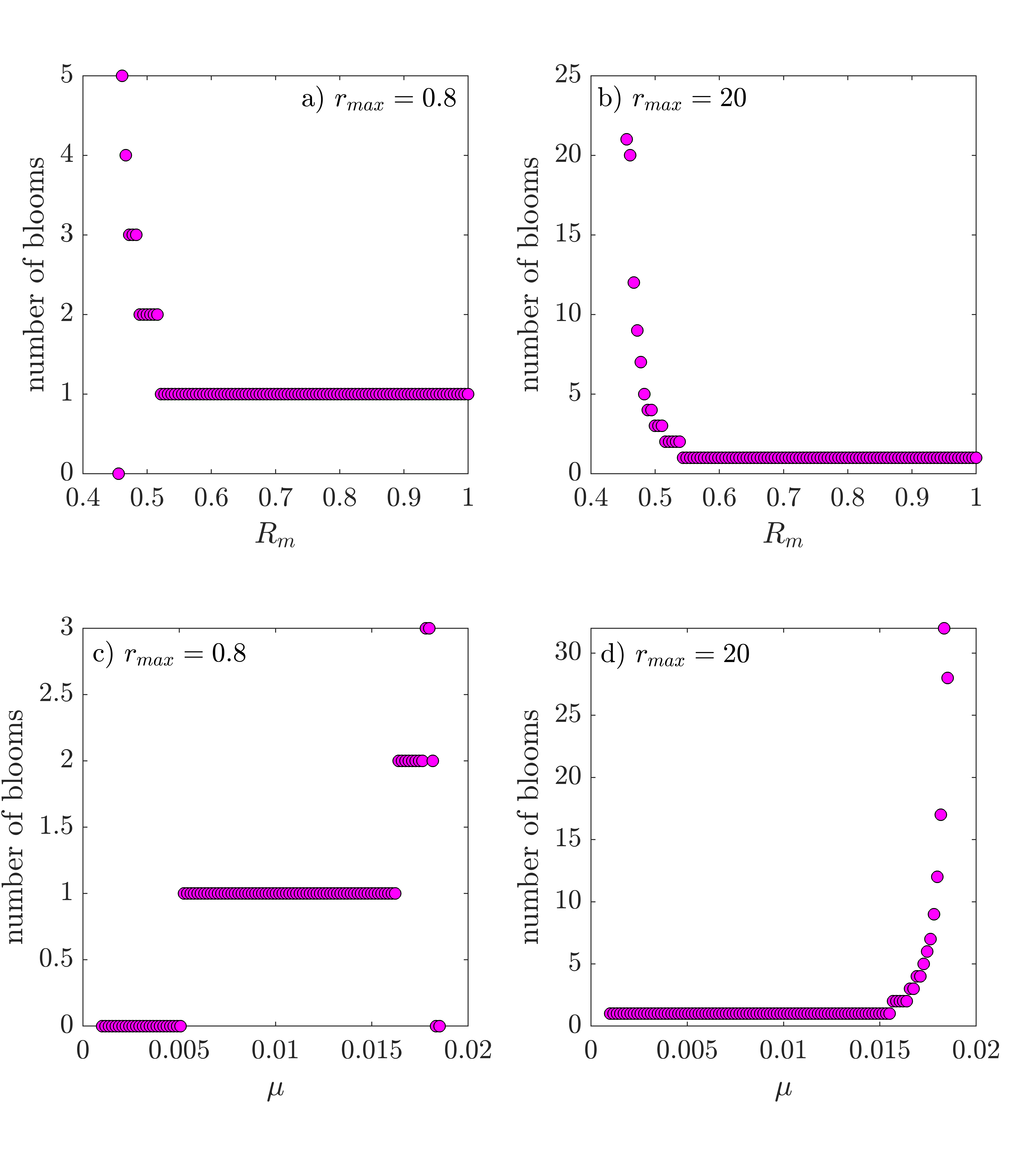}
\caption{The number of recurring blooms $P_{max} > P_b = 60$ depending on the attack rate $R_m$ (a,c) and the mortality rate $\mu$ of the zooplankton (b,d) as well as the maximum growth rate $r_{max}$. Decreasing attack rate $R_m$ and increasing mortality rate $\mu$ promote the occurrence of more and more plankton blooms during the growth rate $r$ increases linearly in time. Parameter: $P(0) = e_P$, $Z(0) = e_Z$; others see Tab. \ref{tab:parameter}.  }
\label{fig:more_spikes}
\end{figure}

The occurrence of more than one plankton bloom while the growth rate $r$ increases in time depends i.a. on the attack rate $R_m$ and the mortality rate $\mu$ of the zooplankton as well as the maximum growth rate $r_{max}$ of the phytoplankton. Of course, $r_{max} = 20$ is far from any realistic approach, nevertheless we choose this value to evaluate how many recurring blooms can be theoretically observed for extreme high maximum growth rates. \\
In the following, we outline our procedure for finding the number of blooms exemplary for $r_{max} = 0.8$ (Fig. \ref{fig:more_spikes}b,d) and different values of the attack rate $R_m$ (Fig. \ref{fig:more_spikes}a). (i) At first, we compute the parameter interval of $R_m$ for which $e(\upsilon t)$ represents the unique stable state. (ii) Then,  we evaluate the rate $\upsilon$ for which the $r$-coordinate of the folded saddle $FS$ is equal to $r_{max}$. (iii) We simulate the three-dimensional TB-model \eqref{app:phy}--\eqref{app:r} for the corresponding attack rates $R_m$ and values of the rate $\upsilon$ ($P(0) = e_{3P}$, $Z(0) = e_{3Z}$ (see Tab. \ref{tab:parameter} for other parameters). (vi) If the maximum phytoplankton density $P_{max}$ exceeds the threshold $P_b = 60$, we increase the parameter $N_b$ which displays the number of blooms, by one. (v) We check randomly if we have detected the correct number of blooms by visible examination of the corresponding trajectory. \\
Obviously, decreasing the attack rate $R_m$ and increasing the mortality rate $\mu$ of the zooplankton cause an increase of the number of recurring blooms (Fig. \ref{fig:more_spikes}). Hence, decreasing the predation pressure on the phytoplankton allows for more and more blooms while $r$ increases linearly in time.

\end{appendices}


\bibliography{literature.bib}


\end{document}